# A Hybrid Vehicle Platoon for Connected and Automated Vehicles: Formulation, Stability Analysis, and Applications

*Yuan Zheng[a], Min Xu[b], Shining Wu[a], Shuaian Wang[a]*

*[a]Department of Logistics & Maritime Studies, The Hong Kong Polytechnic University, Hung Hom, Hong Kong, China*

*[b]Department of Industrial and Systems Engineering, The Hong Kong Polytechnic University, Hung Hom, Hong Kong, China*

## Abstract

Vehicle platooning has the potential to significantly improve traffic throughput and reduce fuel consumption and emissions and thus has attracted extensive attention recently. In this study, we propose a hybrid vehicle platoon system for the connected and automated vehicles (CAVs). First, a hybrid spacing policy combining the constant time gap (CTG) and constant spacing (CS) is formulated for the proposed platoon system, where the leader adopts the CTG and the followers use the CS policy. Based on the $h_2$-norm string stability criteria, the notions of exogenous-head-to-tail string stability and hybrid string stability are newly introduced, and the sufficient conditions of the hybrid string stability in the frequency domain are derived using the Laplace transform. Numerical experiments are conducted to validate the hybrid string stability. Moreover, two typical scenarios and several measurements of effectiveness (MOE) are adopted to verify the effectiveness of the proposed hybrid platoon system in various aspects. The results show that the hybrid platoon system performs better than the CS-based platoon system. It also indicates that the hybrid platoon system has obvious advantages over the CTG-based platoon system under the periodical fluctuation scenario and it is also comparable to the CTG-based platoon system under the large deceleration and acceleration scenario. The findings have demonstrated the merits of the combined implementation of CTG and CS policy in enhancing the performance and applicability of the platoon system for CAVs.

## 1. Introduction

Connected and automated vehicles (CAVs) have received considerable attention due to their potential to drastically improve traffic efficiency, traffic safety, fuel consumption, and emissions (Ma et al., 2017; Shladover, 2018; Talebpour and Mahmassani, 2016; Wu et al., 2021; Yu et al., 2019a; Zheng et al., 2019; Zhou et al., 2019). As a representative application of CAVs, vehicle platoon allows multiple CAVs to operate cooperatively in the form of a platoon (Darbha et al., 2019; Gong et al., 2019; Zhou et al., 2017), thanks to the advanced sensing and vehicle-to-vehicle (V2V) and vehicle-to-infrastructure (V2I) communication technologies. Vehicle platoon enables the CAVs to automatically maintain stable and safe spacings with the predecessors and resolve the traffic disturbances even a small inter-vehicle spacing is adopted (Bian et al., 2019; Wang et al., 2020; Yu et al., 2019b). Recent studies have indicated that the vehicle platoon systems can effectively enhance traffic throughput and flow stability and reduce fuel consumption and emissions with shorter car-following spacing and more exchanged state information (Mahdinia et al., 2020; Xiao et al., 2018; Zhou et al., 2020).

Regarding the vehicle platoon system, local stability and string stability are the most significant properties and should be addressed before practical application (Bian et al., 2019; Feng et al., 2019; Swaroop and Hedrick, 1999; Zhou and Ahn, 2019). The local stability refers to the attenuation ability of a single vehicle against the disturbance from its preceding vehicle. Unlike the local stability, there are various distinct definitions of the string stability to represent the attenuation ability through a string of CAVs under the disturbances, such as $l_2$-norm, $l_p$-norm, $h_2$-norm, and head-to-tail string stability, for spacing error or acceleration (Ge and Orosz, 2014; Naus et al., 2010; Ploeg et al., 2011, 2014; Zhang et al., 2020b; Zhou et al., 2019). Another important component of the vehicle platoon is the car-following spacing policy, which is to determine the desired spacing that a CAV attempts to maintain with respect to the preceding vehicle. The spacing policy plays a significant role since it affects the car-following behavior, stability, and operation performance (e.g., throughput, fuel consumption, and emissions) of platoon systems to a large extent. Specifically, two spacing policies, i.e., constant time gap (CTG) and constant spacing (CS), are commonly used in vehicle platoon systems (Bian et al., 2019; Swaroop and Hedrick, 1999; Xiao et al., 2009; Zheng et al., 2016; Zhou and Ahn, 2019). The CTG policy can improve the string stability of the vehicle platoon but it requires more intensive inter-vehicular communication and will compromise the traffic throughput. On the contrary, the CS policy can potentially lead to higher traffic throughput than the CTG policy. However, it is difficult to maintain the string stability through the whole platoon system, which can adversely affect the operation performance, e.g., fuel consumption and emissions. How to design an effective spacing policy for vehicle platoon systems that achieve better operation performance without sacrificing the stability is important for the implementation of CAVs towards future mobility.

## 1.1. Literature review

Over the past decades, many studies have focused on the development of novel spacing policies and the refinement of existing spacing policies to enhance the stability and operation performance of vehicle platoon systems. The earliest studies for CTG policy focused on the adaptive cruise control dedicated for a single vehicle only (Goñi-Ros et al., 2019). The recent development of V2V/V2I communication has motivated more and more efforts on the cooperative adaptive cruise control for a group of CAVs such as a convoy under the CTG policy (Bian et al., 2019; Darbha et al., 2019; Naus et al., 2010; Ploeg et al., 2011; Zhang et al., 2020a; Zhou and Ahn, 2019), referred to as CTG-based platoon system thereafter. In a CTG-based platoon system, the state information of predecessors can be exchanged and applied to formulate the vehicle controllers through V2V or V2I (Naus et al., 2010; Ploeg et al., 2011; Zhang et al., 2020a) and this formulation can guarantee the string stability of the CTG-based platoon system with a simple predecessor-follower communication topology based on linear feedforward and feedback controllers. Previous experiments and simulations have demonstrated that the CTG-based platoon systems with a variety of feedback and feedforward structures can improve traffic throughput and fuel consumption with guaranteed string stability (Talebpour and Mahmassani, 2016; Wang, 2018; Xiao et al., 2018; Zhang et al., 2020a; Zhou et al., 2019). For example, compared with the predecessor-follower communication topology, the multiple-predecessor-follower communication topology will result in a tighter platoon system, in which a smaller time gap can be adopted to ensure string stability resort to more state information, and thus the traffic throughput of platoon systems is further enhanced (Bian et al., 2019; Darbha et al., 2017; 2019,). Furthermore, the communication delay and sensing delay etc., which are the intrinsic characteristics of the automation and communication systems, can be detrimental to the string stability performance of a CTG-based platoon system since those delays have been found to enlarge the required constant time gap to ensure string stability (Bian et al., 2019; Darbha et al., 2019; Talebpour and Mahmassani, 2016; Wang, 2018; Xiao et al., 2009; Zhou and Ahn, 2019). As such, the traffic throughput is also adversely affected by those delays. To address this problem, many methods have been proposed in the literature (Xing et al., 2020; Zhang et al., 2020a; Zhou and Ahn, 2019), among which the delay compensating method, which is to synchronize the data information collected from the sensing and communication devices and implement the coordinated historical state information of the predecessors to formulate a new car-following spacing policy, was demonstrated to have better and robust performance against the uncertainties of those delays in the platoon system and the potential to achieve better string stability and traffic throughput (Zhang et al., 2020a).

As for CS-based platoon systems, the initial attempt of research aimed to design the controller to regulate a fixed inter-vehicle spacing, using the state information (i.e. position, speed, acceleration) of the preceding vehicle collected by on-board sensor (Seiler et al., 2004; Swaroop and Hedrick, 1999). However, the platoon systems often suffer from the string instability problem. To improve stability, the state information of the leading vehicle is also incorporated into the vehicle controller to maintain the constant spacing. In this way, the CS-based platoon system using a leader-predecessor-follower communication topology can achieve string stability on spacing error (Swaroop and Hedrick, 1999). Some studies further illustrated that the CS-based platoon system can significantly enhance the traffic throughput with the guaranteed string stability (Liu et al., 2001; Xiao et al., 2009; Zhang et al., 2020b). However, analogous to the CTG-based platoon system, the string stability performance of the CS-based platoon system is inevitably affected by the sensing delay and communication delay, etc. (Besselink and Johansson, 2017; Liu et al., 2001; Zhang et al., 2020b). Again, the adverse impacts of delays can be mitigated by the delay compensating method that synchronizes the historical state information of the predecessors. Zhang et al. (2020b) found that the delay compensating method can significantly improve the string stability performance of the CS-based platoon systems. However, most studies investigated the string stability on spacing error only without considering the stability performance on acceleration, which suggests that CS-based platoon systems may have poor performance, e.g., in fuel consumption and emission aspects. Moreover, how the CS-based platoon system effectively follows the exogenous vehicles on roads, especially in complex car-following situations, has not been adequately addressed. Consequently, real-world applications of the CS-based platoon system are rather limited since it is hard to achieve adaptive controls.

As mentioned above, we can see that the CTG and CS policies have distinct advantages and shortcomings. For example, the CTG-based platoon system can achieve improvements in traffic throughput with guaranteed string stability but requires a complex communication topology, and the traffic throughput will decrease with the increase of driving speed since the inter-vehicle distance could become larger. The CS-based platoon system can ensure large and constant traffic throughput but it suffers from string instability and leads to high fuel consumption and emissions. However, all the aforementioned studies focused on either pure CTG-based or pure CS-based platoon systems. To the best of our knowledge, no one has ever formulated and analyzed the performance of a platoon system with a hybrid spacing policy that synergizes the CTG and CS policies and accordingly inherits the benefits of the two policies in one platoon system. In addition, although many studies have examined the operation performance of CTG-based platoon systems, the evaluation of the other platoon systems is limited, such as CS-based platoon systems. Moreover, the performance comparisons among the CTG-based, CS-based, and hybrid platoon systems remain to be explored.

### 1.2 Objectives and contributions

To bridge the above gaps, this study develops a novel hybrid vehicle platoon system that integrates CTG and CS policies in which the leading vehicle follows CTG policy, and the following vehicles are regulated according to CS policy. We introduce a novel notion of exogenous-head-to-tail string stability with respect to the attenuation ability of the last vehicle in the proposed hybrid platoon system to the exogenous leading vehicle and derive the sufficient conditions of string stability through the whole platoon system in the frequency domain by the Laplace transform. Extensive numerical experiments have been conducted to justify the stability of the proposed hybrid platoon system and evaluate its performance in terms of efficiency, stability, energy, emissions, safety, and comfort compared to the CTG-based and CS-based platoon systems. Moreover, the hybrid multi-platoon systems are implemented to test the applicability of the hybrid designs.

The remainder of this study is organized as follows. Section 2 presents the assumptions and problem description. The proposed hybrid platoon system is formulated in Section 3. Section 4 introduces the novel notions of string ability and rigorously demonstrates their validation in hybrid platoon systems. Section 5 elaborates on the experiment settings and the simulation results in detail. Section 6 concludes this paper with future research directions.

## 2. Assumptions and Problem Description

As reviewed in Section 1, we can find that most studies focused on either CS-based or CTG-based platoon systems. Figure 1(a) and Figure 1(b) illustrate a prevalent CS-based platoon system with a leader-predecessor-follower communication topology and a CTG-based platoon system with a two-predecessor-follower communication topology. Let $r$ denote the number of the predecessors with which the subject vehicle can communicate in the communication topology; and obviously we have $r = 2$ in Figure 1. In this section, we propose a novel vehicle platoon system based on a hybrid spacing policy, in which the leading vehicle employs the CTG policy and the following vehicles use the CS policy as illustrated in Figure 1(c). The proposed vehicle platoon system is referred to as the hybrid platoon system throughout the paper. For ease of presentation, the vehicles involved in the platoon system are indexed by $i \in \{0, 1, 2, \ldots, n\}$ in terms of their longitudinal sequence. Specifically, $i = 0$ denotes the exogenous leading vehicle of the hybrid platoon system (See the black vehicle in Figure 1 (c)), $i = 1$ denotes the leading vehicle in the hybrid system that follows the CTG policy, and $i = 2, \ldots, n$ denote the following vehicles in the hybrid system governed by the CS policy.

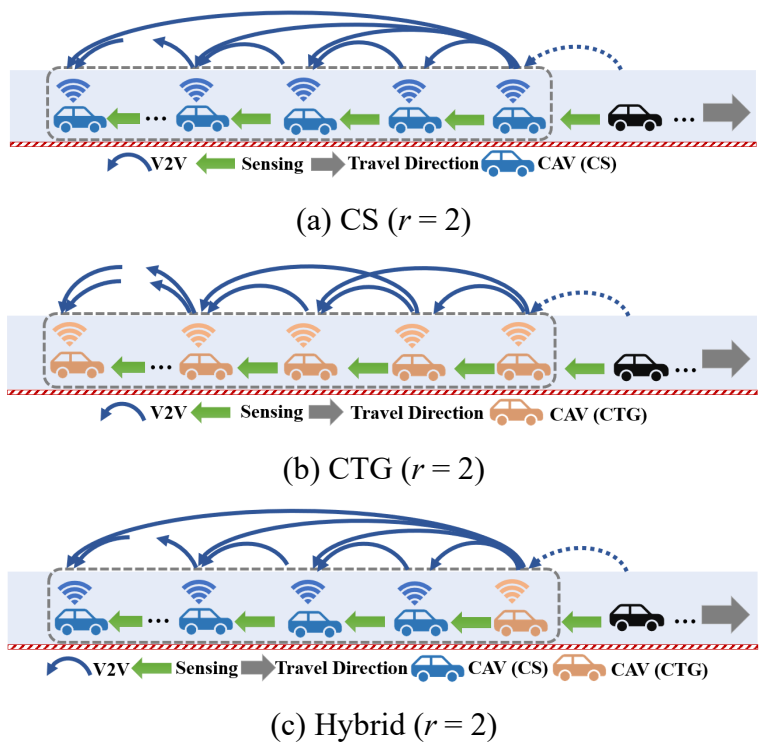

(a) CS ($r = 2$)

(b) CTG ($r = 2$)

(c) Hybrid ($r = 2$)

Figure 1. Illustration of three platoon systems with different spacing policies: (a) CS; (b) CTG; (c) Hybrid.

As for the communication topology, we assume that depending on the type of exogenous leading vehicle $i = 0$, the leading vehicle of the hybrid platoon system, i.e., vehicle $i = 1$, either employs the predecessor-follower communication (if vehicle $i = 0$ is a CAV) or has no communication with the other vehicles (if vehicle $i = 0$ is an automated vehicle (AV) or human-driven vehicle (HDV)). For any subject vehicle $i \in \{1, 2, …, n\}$ in the hybrid platoon system, the position and speed information of its preceding vehicle is perceived by the on-board sensor of the subject vehicle with a sensing delay $\delta_i$, while the acceleration information of the preceding vehicle is exchanged via V2V communication with a communication delay $\theta_i$. In addition to the information of the immediate preceding vehicle, the state information, i.e., position, speed, and acceleration, of all the other predecessors of the subject vehicle will be obtained via V2V communication with distinct communication delays. Moreover, for simplicity, we assume that the following vehicles in the hybrid platoon system, i.e., vehicle $i \in \{2, 3, …, n\}$, have the same control law and controller parameters following the CS policy, whereas the leading vehicle $i = 1$ adopts the control law and controller parameters obeying the CTG policy.

The objective of this study is to mathematically formulate the hybrid platoon system with a hybrid spacing policy, introduce the notations of stability for the new platoon system, and identify the sufficient conditions for its stability by rigorous proofs. Details can be found in the next sections.

## 3. Model Formulation

Following the conversion of literature (Bian et al., 2019; Naus et al., 2010; Ploeg et al., 2011; Zhou and Ahn, 2019), the longitudinal vehicle dynamics for any vehicle $i \in \{1, 2, …, n\}$ in the hybrid platoon system can be formulated by the following linear third-order model

$$\dot{p}_i(t) = v_i(t), \forall i \in \{1, 2, ..., n\} \tag{1}$$

$$\dot{v}_i(t) = a_i(t), \forall i \in \{1, 2, ..., n\} \tag{2}$$

$$\dot{a}_i(t) = -\frac{1}{\varphi}a_i(t) + \frac{1}{\varphi}u_i(t), \forall i \in \{1, 2, ..., n\} \tag{3}$$

where $p_i(t)$, $v_i(t)$, $a_i(t)$ is the front bumper position, speed, and realized acceleration of vehicle $i$ at time $t$ in the hybrid platoon system, respectively, $u_i(t)$ represents the desired acceleration, and $\varphi$ is the actuation time lag to realize the desired acceleration. Since the leading vehicle $i = 1$ and the following vehicles $i \in \{2, 3, …, n\}$ in the hybrid platoon system follow different spacing policies and control laws, we will formulate the spacing policies and control laws of them respectively in the next subsections.

### 3.1. Spacing policy and control law of the leading vehicle

To mitigate the adverse impacts of the delays on controller performance, we assume that the leading vehicle in the hybrid platoon system, i.e. vehicle $i = 1$, follows the delay compensating-based CTG policy proposed by Zhang et al. (2020a), to regulate the inter-distance between vehicles. Therefore, the target equilibrium spacing between the exogenous leader $i = 0$ and the subject vehicle $i = 1$ is formulated as follows:

$$s_1^*(t) = v_1(t)h + \int_{\tau=t-g_1}^{t} v_0(\tau)d\tau + d \tag{4}$$

where $h$ denotes the pre-defined constant time gap which refers to a period during which the front bumper of the subject vehicle approaches the rear bumper of the preceding vehicle at car-following condition, $d$ is the desired inter-vehicle distance in the standstill condition between the front bumper of the following vehicle and the rear bumper of the preceding vehicle, and $g_1$ is the time delay that satisfies $g_1 \geq max(\delta_1, \theta_1)$, which is no less than the sensor delay $\delta_1$ and communication delay $\theta_1$ of the leading vehicle $i = 1$.

The spacing error between the exogenous leader $i = 0$ and the subject vehicle $i = 1$ is given by

$$\begin{aligned}
\Delta s_1(t) &= s_1(t) - s_1^*(t) = \left[ p_0(t) - p_1(t) - L \right] - \left[ v_1(t)h + \int_{\tau=t-g_1}^{t} v_0(\tau)d\tau + d \right] \\
&= p_0(t-g_1) - p_1(t) - L - v_1(t)h - d
\end{aligned} \tag{5}$$

where $s_1(t)$ denotes the actual spacing between the subject vehicle $i = 1$ and its preceding vehicle $i = 0$, i.e., $\Delta s_1(t) = p_0(t) - p_1(t) - L$, and $L$ is the length of the vehicle, which is assumed to be the same for all vehicles.

In the spirit of Bian et al. (2019) and Darbha et al. (2019), the control law of the leading vehicle $i = 1$ is formulated by

$$\begin{aligned}
u_1(t) &= k_s \Delta s_1(t) + k_v \left[ v_0(t-g_1) - v_1(t) \right] + k_a \left[ a_0(t-g_1) - a_1(t) \right] \\
&= k_s \left[ p_0(t-g_1) - p_1(t) - L - v_1(t)h - d \right] + k_v \left[ v_0(t-g_1) - v_1(t) \right] + k_a [a_0(t-g_1) - a_1(t)]
\end{aligned} \tag{6}$$

where $k_s$, $k_v$, and $k_a$ denote the controller parameters of the subject vehicle $i = 1$, which are the coefficients of the spacing error, speed difference, and acceleration difference acquired via V2V communication, respectively. Note that the acceleration information of the preceding vehicle is not available if the preceding vehicle, i.e., vehicle $i = 0$, is an AV or an HDV.

### 3.2. Spacing policy and control law of the following vehicles

For the following vehicles in the hybrid platoon system, i.e., vehicle $i \in \{2, 3, \ldots, n\}$, the delay compensating-based CS policy is applied with a leader-predecessor-follower communication topology. Therefore, the target spacing is calculated by

$$s_i^*(t) = d + \int_{\tau=t-g_i}^{t} v_{i-1}(\tau)d\tau, \forall i \in \{2,3,...,n\} \tag{7}$$

where $g_i$ is the time delay no smaller than the sensor delay $\delta_i$, communication delay $\theta_i$, and the difference of the communication delay $\Delta\theta_{1,i}$ between two consecutive vehicles relative to the leading vehicle in the hybrid platoon system, namely, $g_i \geq max(\delta_i, \theta_i, \Delta\theta_{1,i})$. Note that $\Delta\theta_{1,i}$ is calculated by $\Delta\theta_{1,i} = \theta_{1,i} - \theta_{1,i-1}$, where $\theta_{1,i}$ represents the communication delay between the leading vehicle $i = 1$ and the subject vehicle $i$.

The spacing error between the preceding vehicle ($i$-1) and the subject vehicle $i$ is calculated by

$$\begin{aligned}\Delta s_i(t) &= s_i(t) - s_i^*(t) = \left[ p_{i-1}(t) - p_i(t) - L \right] - \left[ d + \int_{\tau=t-g_i}^{t} v_{i-1}(\tau)d\tau \right] \\ &= \left[ p_{i-1}(t) - \int_{\tau=t-g_i}^{t} v_{i-1}(\tau)d\tau \right] - p_i(t) - L - d \qquad , \forall i \in \{2,3,...,n\} \\ &= p_{i-1}(t-g_i) - p_i(t) - L - d \end{aligned} \tag{8}$$

and the spacing error relative to the leading vehicle $i = 1$ is given by

$$\begin{aligned}\Delta s_{1,i}(t) &= \sum_{k=2}^{i} \Delta s_k(t + \sigma_k - \sigma_i) \\ &= \sum_{k=2}^{i} \left[ p_{k-1}(t + \sigma_k - \sigma_i - g_k) - p_k(t + \sigma_k - \sigma_i) - L - d \right], \forall i \in \{2,3,...,n\} \\ &= p_1(t - \sigma_i) - p_i(t) - \sum_{k=2}^{i}(L + d) \end{aligned} \tag{9}$$

where $\sigma_i$ is the accumulated time delay along the hybrid platoon system, namely, $\sigma_i = \sum_{k=2}^{i} g_k$.

On basis of the CS policy, the control law is formulated as follows aiming to control the subject vehicle to track the position of both the preceding vehicle with desired distance and the leading vehicle with desired accumulated distance (Liu et al., 2001; Swaroop and Hedrick, 1999):

$$
\begin{aligned}
u_i(t) &= \frac{1}{1+q_3}[a_{i-1}(t-g_i)+q_3 a_1(t-\sigma_i)+(q_1+\lambda)\Delta\dot{s}_i(t)+q_1\lambda\Delta s_i(t)+(q_4+\lambda q_3)\Delta\dot{s}_{1,i}(t)+\lambda q_4\Delta s_{1,i}(t)] \\
&= \frac{1}{1+q_3}\left\{a_{i-1}(t-g_i)+q_3 a_1(t-\sigma_i)+(q_1+\lambda)\left[v_{i-1}(t-g_i)-v_i(t)\right]+q_1\lambda(p_{i-1}(t-g_i)-p_i(t)-d)\right\} \\
&\quad + \frac{1}{1+q_3}\left\{(q_4+\lambda q_3)\left[v_1(t-\sigma_i)-v_i(t)\right]+\lambda q_4\left[p_1(t-\sigma_i)-p_i(t)-\sum_{k=2}^{i}(L+d)\right]\right\}, \forall i\in\{2,3,...,n\}
\end{aligned}
\tag{10}
$$

where $q_1$, $q_3$, $q_4$, and $\lambda$ denote the controller parameters of the following vehicles $i\in\{2, 3, …, n\}$ in the proposed hybrid system. It can be found that the control law of the subject vehicle is dependent on the acceleration of both the leading vehicle and the preceding vehicle, and the speed difference and spacing error of the subject vehicle with respect to both the leading vehicle and the preceding vehicle.

## 4. Stability Analysis

Local stability and string stability are the critical properties serving the vehicle platoon systems. In what follows, we will derive the sufficient conditions for the local stability on spacing error and acceleration, the string stability on spacing error, and introduce a novel string stability on acceleration for the hybrid platoon system.

### 4.1. Local stability analysis

In the hybrid platoon system, the local stability conditions for each vehicle governed either by CS or CTG policy in the hybrid platoon system can be readily derived as follows based on previous studies (Bian et al., 2019; Zhang et al., 2020b; Zhou and Ahn, 2019).

**Definition 1** (Zhou et al., 2020): A platoon system is linearly local stable if and only if all eigenvalues of the characteristic equation in the closed-loop platoon system has strictly negative real parts.

**Proposition 1** (Zhang et al., 2020b): the hybrid platoon system is local stable on spacing error if the following conditions are satisfied:

$$\begin{cases} 1 + q_3 > 0 \\ (q_1 + q_4)\lambda > 0 \\ \lambda(1 + q_3) > (\lambda\varphi - 1)(q_1 + q_4) \end{cases} \tag{11}$$

**Proposition 2** (Bian et al., 2019): the hybrid platoon system is local stable on acceleration if the following conditions are satisfied:

$$\begin{cases} 1 + k_a > 0 \\ k_p > 0 \\ k_v + hk_p > 0 \\ (1 + k_a)(k_v + hk_p) - \varphi k_p > 0 \end{cases} \tag{12}$$

### *4.2. String stability analysis*

This subsection analyses the string stability of the hybrid platoon systems, particularly the $h_2$-norm string stability, both on spacing error and acceleration. We will introduce the definitions of the string stability for the hybrid platoon systems and derive the sufficient conditions by rigorous mathematical proofs.

### *4.2.1 String stability on spacing error*

**Definition 2** (Zhang et al., 2020b): A platoon system is $h_2$-norm string stable on spacing error if and only if

$$\frac{\|\Delta s_i(z)\|_2}{\|\Delta s_{i-1}(z)\|_2} \leq 1, \forall w > 0, z = jw, \forall i \in \{2, 3, ..., n\} \tag{13}$$

where $\|.\|$ denotes the $h_2$ norm, $\Delta s_i(z)$ is the Laplace transform of the spacing error of $\Delta s_i(t)$, $j$ is the imaginary unit, and $w$ is the frequency.

**Proposition 3** (Zhang et al., 2020b): the hybrid platoon system is string stable on spacing error if the following conditions are satisfied:

$$\left\| \frac{z^2 + (q_1 + \lambda)z + q_1\lambda}{(1 + q_3)z^2(1 + z\varphi) + (q_1 + \lambda + q_4 + q_3\lambda)z + (q_1 + q_4)\lambda} \right\| \leq 1, \forall w > 0, z = jw \tag{14}$$

*4.2.2 String stability on acceleration*

**Definition 3** (Zhou and Ahn, 2019): A platoon system is $h_2$-norm string stable on acceleration if and only if

$$\frac{\left\|a_i(z)\right\|_2}{\left\|a_{i-1}(z)\right\|_2} \leq 1, \forall w > 0, z = jw, \forall i \in \{2,3,...,n\} \tag{15}$$

where $a_i(z)$ denotes the Laplace transform of the acceleration $a_i(t)$.

We can see from Eq. (15) that the string stability criteria are strict since they require disturbance attenuation for every vehicle in a platoon. In fact, Ge and Orosz (2014) have found that it is hard to achieve the string stability on acceleration for the platoon systems in a connected cruise control way with complex communication topologies whose control law additionally depends on the acceleration information of multiple predecessors. Therefore, they proposed the notion of head-to-tail string stability on acceleration and derived the sufficient and necessary stability conditions for the proposed platoon systems. The head-to-tail string stability on acceleration denotes the attenuation of acceleration of the last vehicle with respect to the first vehicle in the platoon system as follows:

**Definition 4** (Ge and Orosz, 2014): A platoon system is $h_2$-norm head-to-tail string stable on acceleration if and only if

$$\frac{\left\|a_n(z)\right\|_2}{\left\|a_1(z)\right\|_2} \leq 1, \forall w > 0, z = jw \tag{16}$$

where $a_1(z)$ and $a_n(z)$ is the Laplace transform of the acceleration of the leading vehicle $a_1(t)$ and the last vehicle $a_n(t)$ in the platoon system, respectively.

It can be seen that the head-to-tail string stability criteria on acceleration in Eq. (16) are weaker than the string stability criteria on acceleration in Eq. (15). However, we find that the hybrid platoon systems cannot ensure the head-to-tail string stability on acceleration as demonstrated in Appendix B and the same applies for the CS-based platoon systems. Nevertheless, unlike the CS-based platoon system, the leader in the hybrid platoon system follows the CTG policy, which is likely to assist the platoon system to attenuate the disturbances from the exogenous leader. In light of this, we introduce a novel notion of string stability for the proposed hybrid platoon systems as follows:

**Definition 5**: A platoon system is $h_2$-norm exogenous-head-to-tail string stable if and only if

$$\frac{\|a_n(z)\|_2}{\|a_0(z)\|_2} \leq 1, \forall w > 0, z = jw \tag{17}$$

where $a_0(z)$ is the Laplace transform of the exogenous leader's acceleration $a_0(t)$ of the platoon system.

The above stability definition is to guarantee that the acceleration of the last vehicle in the platoon system is smaller than that of the exogenous leader of the platoon system. Compared with the head-to-tail string stability concerning the attenuation ability of the last vehicle to the leading vehicle through the platoon, the exogenous-head-to-tail string stability indicates the attenuation ability of the last vehicle in the platoon to the exogenous leader. It implies the attenuation ability of the whole platoon system against exogenous disturbances, which can be regarded as a novel string stability on acceleration. Note that the exogenous-head-to-tail string stable (stability) is referred to as ex-head-to-tail string stable (stability) in the rest of the paper. In the following proposition, we will derive the sufficient conditions for the ex-head-to-tail string stability on acceleration for the hybrid platoon systems.

**Proposition 4:** the hybrid platoon system is ex-head-to-tail string stable on acceleration if the following conditions are satisfied:

$$\left\| \frac{1 + G(z)K(z) + G(z)F(z)h}{G(z)K(z)} \right\| \geq max(\frac{\|C(z)\|}{\|A(z)\| - \|B(z)\|}, 1), \forall w > 0, z = jw \tag{18}$$

where $G(z)$, $K(z)$, and $F(z)$ denotes the Laplace transform of the vehicle dynamic, controller terms (except the time gap), and time gap term in the control law of the leader in the platoon, respectively, and $A(z)$, $B(z)$, and $C(z)$ denote the Laplace transforms of controller terms related to the vehicles' longitudinal positions in the control law of the following vehicles in the platoon.

**Proof.** We will first prove that the following inequalities hold if Eq. (18) is satisfied:

$$\frac{\|a_1(z)\|}{\|a_0(z)\|} \leq 1, \forall w > 0, z = jw \tag{19}$$

Specifically, we will examine the leading vehicle, i.e., vehicle $i = 1$. This vehicle follows the CTG policy in the hybrid platoon system. The relation of the acceleration between the vehicle $i = 1$ and the exogenous leader $i = 0$ can be expressed as follows based on Eq. (6) using the Laplace transform:

$$\frac{a_1(z)}{a_0(z)} = \frac{G(z)K(z)}{1 + G(z)K(z) + G(z)F(z)h} e^{-g_1 z} \qquad (20)$$

with

$$\begin{cases} G(z) = 1/(z^2 + z^3 \varphi) \\ K(z) = k_a z^2 + k_v z + k_p \\ F(z) = k_p z \end{cases} \qquad (21)$$

It follows from Eq. (20) that

$$\begin{aligned} \frac{\|a_1(z)\|}{\|a_0(z)\|} &\leq \left\| \frac{G(z)K(z)}{1 + G(z)K(z) + G(z)F(z)h} \right\| \left\| e^{-g_1 z} \right\| \\ &\leq \left\| \frac{G(z)K(z)}{1 + G(z)K(z) + G(z)F(z)h} \right\| \leq 1, \forall w > 0, z = jw \end{aligned} \qquad (22)$$

where the last inequality follows from Eq. (18). Therefore, we have proved that Eq. (19) is satisfied under Condition (18).

As for the vehicles $i \in \{2, 3, \ldots, n\}$, Eq. (10) in the frequency domain using the Laplace transform can be written as follows：

$$A(z)p_i(z) = B(z)p_{i-1}(z)e^{-g_i z} + C(z)p_1 e^{-\sigma_i z}, \forall i \in \{2,3,...,n\} \qquad (23)$$

with

$$\begin{cases} A(z) = (1+q_3)z^2(1+z\varphi) + (q_1 + \lambda + q_4 + q_3\lambda)z + (q_1 + q_4)\lambda \\ B(z) = z^2 + (q_1 + \lambda)z + q_1\lambda \\ C(z) = q_3 z^2 + (q_4 + q_3\lambda)z + q_4\lambda \end{cases} \qquad (24)$$

where $p_i$(z), $p_{i-1}$(z), and $p_1$(z) is the Laplace transform of the front bumper position of vehicle $i$ (i.e., $p_i(t)$), vehicle ($i$-1) (i.e., $p_{i-1}(t)$), and the leading vehicle $i = 1$ (i.e., $p_1(t)$), in the platoon system, respectively.

Next, we will use mathematical induction to prove that if Eqs. (18) are satisfied, the following inequalities hold:

$$\frac{\|p_i(z)\|}{\|p_1(z)\|} \leq \left\|\frac{1+G(z)K(z)+G(z)F(z)h}{G(z)K(z)}\right\|, \forall w>0, z=jw, i \in \{1,2,3,...,n\} \tag{25}$$

Particularly, for $i=1$, we have

$$\frac{\|p_1(z)\|}{\|p_1(z)\|} = 1 \leq \left\|\frac{1+G(z)K(z)+G(z)F(z)h}{G(z)K(z)}\right\|, \forall w>0, z=jw \tag{26}$$

Suppose that the following inequalities hold for vehicle ($i$-1), $\forall\, i \in \{2, 3, ..., n\}$.

$$\frac{\|p_{i-1}(z)\|}{\|p_1(z)\|} \leq \left\|\frac{1+G(z)K(z)+G(z)F(z)h}{G(z)K(z)}\right\|, \forall w>0, z=jw \tag{27}$$

It follows from Eq. (18) that

$$\left\|\frac{1+G(z)K(z)+G(z)F(z)h}{G(z)K(z)}\right\| \geq \frac{\|C(z)\|}{\|A(z)\|-\|B(z)\|}, \forall w>0, z=jw \tag{28}$$

By simple manipulation, we can further obtain

$$\frac{\|C(z)\|}{\|A(z)\|} \leq (1-\frac{\|B(z)\|}{\|A(z)\|})\left\|\frac{1+G(z)K(z)+G(z)F(z)h}{G(z)K(z)}\right\|, \forall w>0, z=jw \tag{29}$$

Therefore, we have

$$\begin{aligned}
\frac{\|p_i(z)\|}{\|p_1(z)\|} &= \left\|\frac{B(z)p_{i-1}(z)e^{-g_iz}}{A(z)p_1(z)} + \frac{C(z)p_1e^{-\sigma_iz}}{A(z)p_1(z)}\right\| \\
&\leq \frac{\|B(z)\|}{\|A(z)\|}\frac{\|p_{i-1}(z)\|}{\|p_1(z)\|}\|e^{-g_iz}\| + \frac{\|C(z)\|}{\|A(z)\|}\|e^{-\sigma_iz}\| \leq \frac{\|B(z)\|}{\|A(z)\|}\frac{\|p_{i-1}(z)\|}{\|p_1(z)\|} + \frac{\|C(z)\|}{\|A(z)\|} \\
&\leq \frac{\|B(z)\|}{\|A(z)\|}\left\|\frac{1+G(z)K(z)+G(z)F(z)h}{G(z)K(z)}\right\| + (1-\frac{\|B(z)\|}{\|A(z)\|})\left\|\frac{1+G(z)K(z)+G(z)F(z)h}{G(z)K(z)}\right\| \\
&= \left\|\frac{1+G(z)K(z)+G(z)F(z)h}{G(z)K(z)}\right\|, \forall i \in \{2,3,...,n\}
\end{aligned} \tag{30}$$

where the first equality follows from Eq. (23) and the third inequality follows from Eq. (27) (for the first term) and Eq. (29) (for the second term).

The above recursive principal indicates that $\dfrac{\|p_i(z)\|}{\|p_1(z)\|} \leq \left\|\dfrac{1+G(z)K(z)+G(z)F(z)h}{G(z)K(z)}\right\|$ always holds for any vehicle $i \in \{2, 3, ..., n\}$.

According to the property of Laplace transform, we have $\dfrac{\|p_i(z)\|_2}{\|p_1(z)\|_2} = \dfrac{\|a_i(z)\|_2}{\|a_1(z)\|_2}$. Hence, based on Eqs. (22) and (25), we can prove as follows that Eq. (17) in Definition 5 is satisfied for the hybrid platoon system:

$$\frac{\|a_n(z)\|}{\|a_0(z)\|} = \frac{\|a_n(z)\|}{\|a_1(z)\|}\frac{\|a_1(z)\|}{\|a_0(z)\|} \leq \left\|\frac{1+G(z)K(z)+G(z)F(z)h}{G(z)K(z)}\right\| \times \left\|\frac{G(z)K(z)}{1+G(z)K(z)+G(z)F(z)h}\right\| = 1 \quad (31)$$

This completes the proof of Proposition 4. □

We further define the hybrid string stability for a platoon system and summarize the sufficient conditions for hybrid string stability of the proposed hybrid platoon system as follows.

**Definition 6:** A hybrid platoon system is hybrid string stable if and only if Eqs. (13) and (17) are satisfied.

The above stability definition is to guarantee that the spacing error can attenuate in the platoon and the acceleration of the last vehicle in the platoon is also no larger than the exogenous leader of the platoon. The notion emphasizes the attenuation ability of the whole platoon system on both acceleration and spacing error against the disturbances of exogenous leaders. In this way, it can enhance the string stability performance and has the potential to improve the operation performance of the platoon system.

**Proposition 5:** A hybrid platoon system is hybrid string stable if Eqs. (14) and (18) are satisfied.

## 5. Numerical Experiments

In this section, numerical experiments are conducted to verify the effectiveness of the proposed hybrid platoon system from various aspects in comparison with the CTG-based and CS-based platoon systems. We will first present two instances to validate the mathematical proof of stability analysis. The performance of the three types of platoon systems will be compared extensively. Moreover, the performance comparison will be extended to three types of multi-platoon systems. Finally, we summarize the rankings of different platoon systems in several performance aspects. The parameter settings of the platoon systems for numerical experiments are illustrated in Table 1 according to the previous studies (Bian et al., 2019; Darbha et al., 2019; Xiao et al., 2009; Zhang et al., 2020b).

Table 1. Parameter settings for numerical experiments

| Parameter | Notation | Value |
|---|---|---|
| Simulation time | $T$ | 120 s |
| Time step | $\Delta t$ | 0.1 s |
| Vehicle length | $L$ | 5 m |
| Desired standstill distance | $d$ | 5 m |
| Actuation time lag | $\varphi$ | 0.5 s |
| Sensor delay | $\delta$ | 0.02 s |
| Communication delay | $\theta$ | 0.1 s |
| Communication delay of vehicle $i$ to the leading vehicle | $\theta_{1,i}$ | $0.1i$ s |

### 5.1. Stability verification

In this subsection, numerical experiments will be conducted to illustrate the stability analysis. Particularly, we will present two instances of the hybrid platoon system with hybrid string stability (i.e., the hybrid stability conditions are fully satisfied) and hybrid string instability (i.e., the hybrid stability conditions are not fully satisfied). Both platoon instances are composed of five CAVs with the same constant time gap $h$ set at 1.4 s and the same controller parameters set as follows: $q_1 = 0.4$, $q_3 = 0.9$, $q_4 = 0.6$, $k_s = 0.1$, $k_v = 0.7$, $k_a = 0.84$ except that $\lambda = 0.1$ in the former instance while $\lambda = 0.3$ in the latter instance. It can be verified that the parameters in the former instance satisfy the stability conditions in Proposition 5, suggesting that these parameters are located in the feasible region of the string stability on spacing error and the ex-head-to-tail string stability on acceleration, whereas the parameters in the latter instance satisfy all conditions in Proposition 5 except for Eq. (18), indicating that the parameters in the latter instance are within the feasible region of the string stability on spacing error but outside of the feasible region of ex-head-to-tail string stability on acceleration.

Recall that by Definition 6, a hybrid platoon system is hybrid string stable if both the two norms in Eqs. (13) and (17) are smaller than 1 for any $w$; otherwise, it is string unstable. We visualize the norm of $e_i/e_{i-1}$ in Eq. (13) and the norm of $a_5/a_0$ under frequency $w$ in Eq. (17) in the above two instances in Figure 2(a) and (b), respectively. As expected, it shows that the hybrid platoon system is hybrid string stable in the former instance, i.e., string stable on spacing error and ex-head-to-tail string stable on acceleration, since both norms are less than 1 for any $w$, whereas the platoon system becomes hybrid string unstable in the latter instance because the condition in Proposition 5, i.e., Eq. (18), is not satisfied.

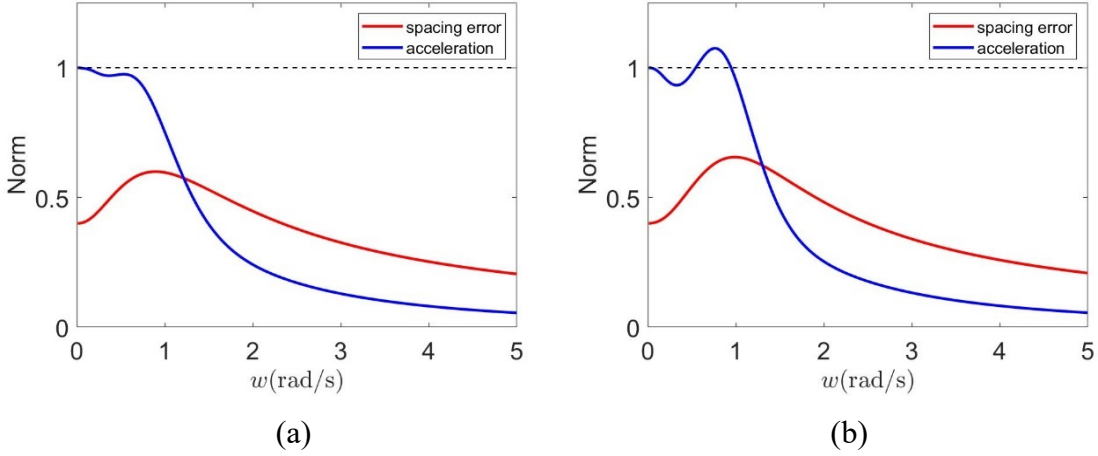

Figure 2. Norms of $e_i/e_{i-1}$ and $a_5/a_0$ in the hybrid platoon system: (a) hybrid string stability; (b) hybrid string instability.

We further illustrate the temporal spacing error and acceleration of the above two instances by simulation experiments in Figure 3. It is assumed that the exogenous leader of the hybrid system performs the periodical acceleration and deceleration ranging from 2.3 m/s² and -2.3 m/s² for 80 s. For the former instance, as shown in Figure 3(a), the spacing error decreases or converges to zero when the disturbance propagates to the tail vehicle in the platoon system. It is found that the leading vehicle using the CTG policy in the platoon has an opposite phase of spacing error compared with the followers. This is because the magnitude of its deceleration is larger than that of actual spacing when the disturbance propagates from the exogenous leader to the leading vehicle. It suggests that the target equilibrium spacing of the leading vehicle is smaller than the actual spacing relative to the exogenous leader. Therefore, there exists some extra space resulting in the opposite phase of the spacing error. In addition, we can observe from Figure 3(b) that the tail vehicle in the hybrid system has a smaller acceleration than the exogenous leader, even if the magnitude of the acceleration of the followers gradually becomes larger. It has demonstrated that the exogenous disturbance can attenuate through the whole platoon system and the hybrid platoon system with the stability parameters can guarantee the convergence of spacing error and the gradual attenuation of the acceleration at a platoon level. On the contrary, for the latter instance, although the spacing error converges in Figure 3(c), the acceleration of the tail vehicle in the platoon system is amplified relative to that of the exogenous leader, as shown in Figure 3(d). In conclusion, the results of the numerical experiments are in accordance with the theoretical analysis in Section 4.

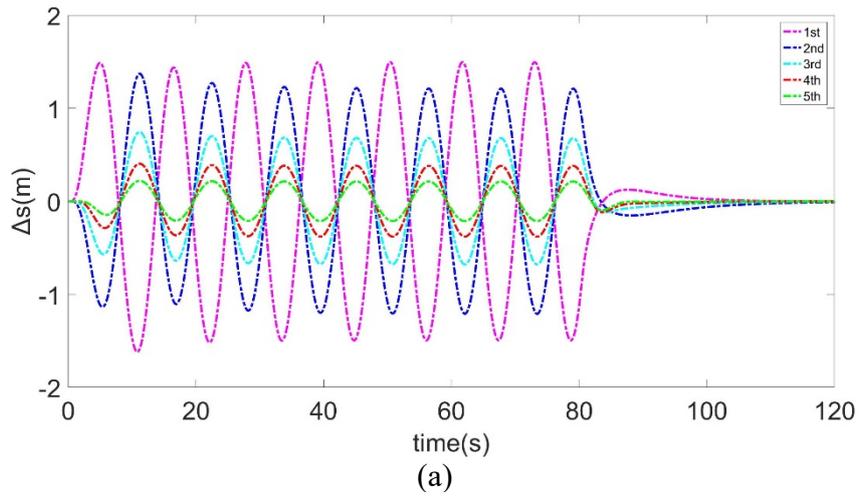

(a)

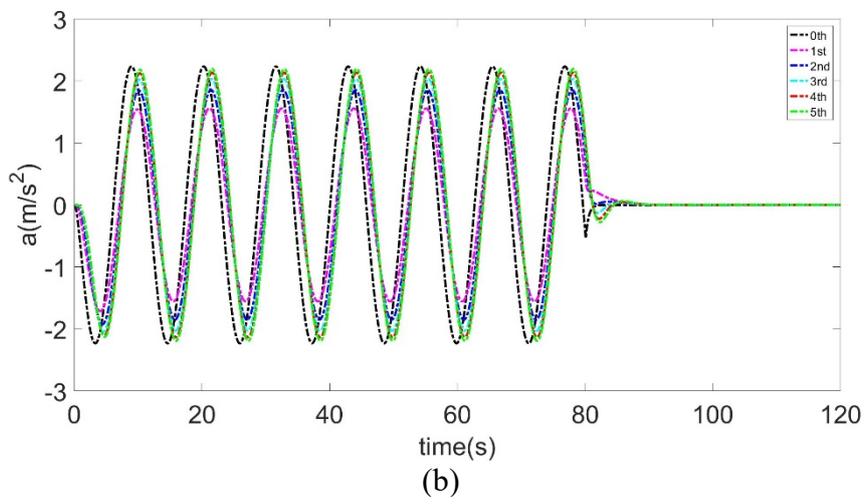

(b)

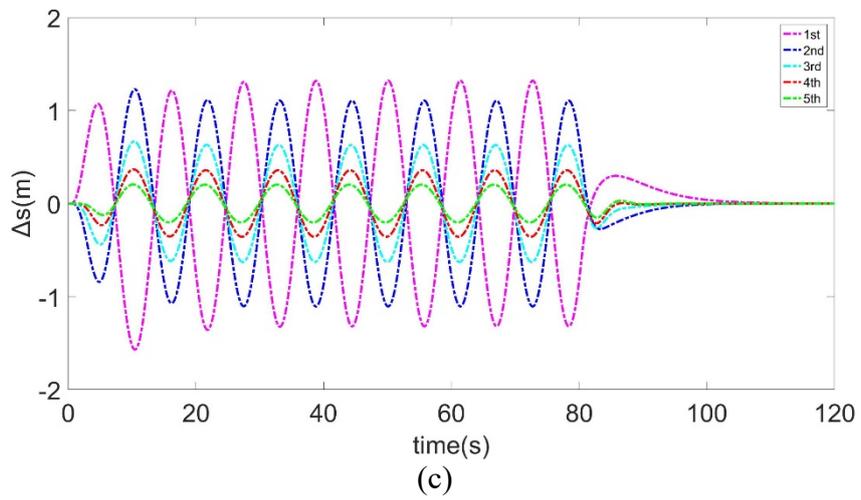

(c)

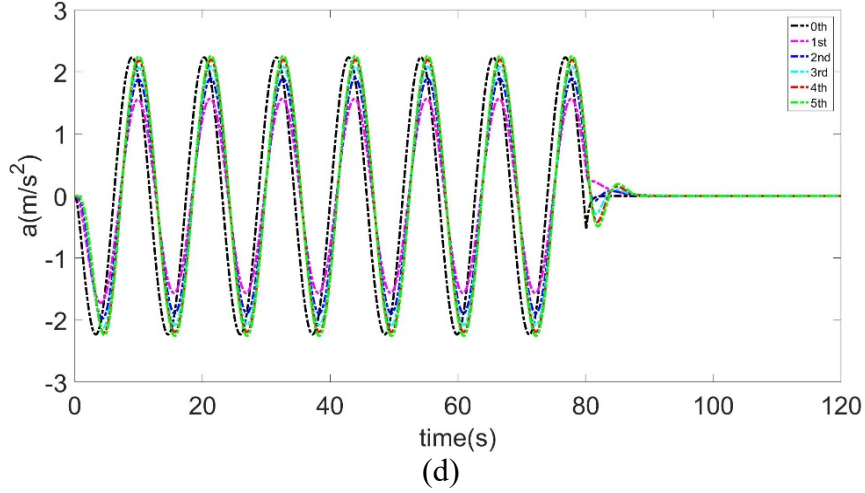

(d)

Figure 3. Motion profile of the hybrid platoon system: (a) spacing error with hybrid string stability; (b) acceleration with hybrid string stability; (c) spacing error with hybrid string instability; (d) acceleration with hybrid string instability.

### *5.2. Comparisons of different platoon systems*

In this subsection, we will compare the performance of three types of platoon systems, namely, CS-based, CTG-based, and hybrid platoon systems, in terms of efficiency, stability, energy, emissions, safety, and comfort. As the communication topology will also influence the string and operation performance, for the CS-based and hybrid platoon systems, we consider the most popular implementation with leader-predecessor-follower communication topology, i.e., $r = 2$ (see Figure 1 (a)), while for the CTG-based system, both two-predecessor-follower communication topology ($r = 2$) and three-predecessor-follower communication topology ($r = 3$) are prominent and will thus be presented. The parameters for the three types of platoon systems are set as follows (Bian et al., 2019; Darbha et al., 2019; Zhang et al., 2020b):

- **CS-based platoon system:** $\lambda = 0.10$, $q_1 = 0.40$, $q_3 = 0.90$, and $q_4 = 0.60$.
- **CTG-based platoon system:** $k_s = 0.10$, $k_v = 1.67$, $k_a = 0.84$, $h_{leader} = 0.594$ s, $h_{follower} = 0.262$ s ($r = 2$), and $h_{follower} = 0.198$ s ($r = 3$).
- **Hybrid platoon system:** $\lambda = 0.1$, $q_1 = 0.4$, $q_3 = 0.9$, $q_4 = 0.6$, $k_s = 0.1$, $k_v = 0.7$, $k_a = 0.84$, and $h_{leader} = 1.40$ s.

Several measurements of effectiveness (MOE) in previous literature including the traffic outflow, dampening ratio, fuel consumption, emission measurements, modified time-to-collision, and maximum jerk will be adopted to evaluate the effectiveness of the proposed platoon system in the aforementioned aspects. The following two disturbance scenarios with a CAV as the exogenous leading vehicle of three types of platoon systems are used for the tests (Cui et al., 2021; Zhang et al., 2020a; Zheng et al., 2020b):

- **Scenario 1-Periodic Fluctuation**: The exogenous leader of the platoon system performs periodic acceleration and deceleration. The disturbance period lasts for 80 s with the acceleration ranging from -2.3 m/s$^2$ and 2.3 m/s$^2$.
- **Scenario 2-Large Deceleration and Acceleration**: The exogenous leader moves at a constant speed 30 m/s for 30 s and then decelerates to 10 m/s with a constant deceleration of -2.5 m/s$^2$. Then the vehicle maintains the driving speed for another 30 s and restores to the original speed with a constant acceleration of 2.5 m/s$^2$.

*5.2.1. Efficiency*

In this subsection, we will compare the efficiency of different platoon systems in terms of traffic outflow by simulations under the above two scenarios. The traffic outflow $Q$ (veh/s) is defined as the ratio of the total number of vehicles $n$ to the time interval between the first vehicle $t_1$ and the last vehicle $t_n$ at the downstream measurement position as follows (Duret et al., 2020):

$$Q = n/(t_n - t_1) \tag{32}$$

To explore the impact of the platoon size (i.e., the number of CAVs in a platoon system), we simulate the platoon systems with the number of vehicles ranging from 4 to 10 and report the traffic outflow of the platoon systems for each specific number of CAVs, e.g., '$n = 4$', and the average traffic outflow of all platoon systems with different sizes for each type of platoon, i.e., 'Average', in Table 2. It shows that the hybrid platoon system using CTG and CS policy has better performance in traffic efficiency than the CTG-based one as it has a larger average traffic outflow. However, both of them are not as good as the CS-based platoon system under Scenario 1. Moreover, we find that for the CTG-based and hybrid systems, the traffic outflows become larger as the number of CAVs increases, implying that the increment of CAV members is helpful in improving the traffic throughput, especially for the hybrid system. The traffic outflow of the hybrid system is greater than the CTG-based system when it incorporates 6 or more CAVs under Scenario 1. Although the CS-based platoon system has the largest traffic flow among all platoon systems, the traffic outflow is not always enlarged with the increments in number of CAVs mainly due to the string instability. Generally speaking, the hybrid system is superior to the CTG-based system but weaker than the CS-based system in terms of traffic efficiency. Under Scenario 2, the traffic outflows of all platoon systems become larger and similar findings can be observed for the comparisons of the three platoon systems. In summary, the results have demonstrated the advantages of the CS policy on maximizing traffic throughput, followed by the hybrid and CTG-based systems in order.

Table 2. Traffic outflow of different platoon systems under Scenarios 1 and 2

| Scenario | System | $r$ | $n=4$ | $n=5$ | $n=6$ | $n=7$ | $n=8$ | $n=9$ | $n=10$ | Average |
|----------|--------|-----|-------|-------|-------|-------|-------|-------|--------|---------|
| Scenario 1 | CS-based | 2 | 2.22 | 2.27 | 2.30 | 2.33 | 2.28 | 2.30 | 2.32 | 2.28 |
| | CTG-based | 2 | 1.29 | 1.31 | 1.33 | 1.34 | 1.35 | 1.36 | 1.36 | 1.33 |
| | | 3 | 1.38 | 1.38 | 1.42 | 1.45 | 1.45 | 1.47 | 1.49 | 1.43 |
| | Hybrid | 2 | 1.29 | 1.38 | 1.50 | 1.59 | 1.63 | 1.69 | 1.75 | 1.54 |
| Scenario 2 | CS-based | 2 | 3.07 | 3.12 | 3.00 | 3.04 | 3.07 | 3.00 | 3.03 | 3.04 |
| | CTG-based | 2 | 1.45 | 1.49 | 1.54 | 1.55 | 1.57 | 1.58 | 1.59 | 1.53 |
| | | 3 | 1.60 | 1.67 | 1.67 | 1.71 | 1.74 | 1.74 | 1.75 | 1.69 |
| | Hybrid | 2 | 1.48 | 1.64 | 1.76 | 1.89 | 1.97 | 2.05 | 2.13 | 1.84 |

*5.2.2. Stability*

In this subsection, we will evaluate the stability performance against the exogenous disturbances for the three platoon systems. In particular, the dampening ratio (DR) is used to measure the stability performance of the platoon system (Zhou et al., 2020) as shown below:

$$DR = \frac{\|a_n(z)\|_2}{\|a_0(z)\|_2} \tag{33}$$

The aforementioned two disturbance scenarios will be employed in the simulations. Likewise, both the DR and the average DR of all platoon systems with different sizes for each type of platoon system are reported in Table 3. It can be found that the DR and average DR are smaller than 1 for the hybrid and CTG-based platoon systems, while they are larger than 1 for the CS-based platoon system under Scenario 1. The results suggest that compared with the CS-based system, the hybrid and CTG-based system have better stability performance that makes the disturbances attenuate from the exogenous leader to the last vehicle in the system. The advantage of the hybrid systems over CS-based system should be attributed to the relatively large time gap of leader in the hybrid system, which essentially acts as a buffer against periodical disturbances from the exogenous leader. As a result, the hybrid formulation shows a similar stability performance to the CTG-based system with commonly-accepted good stability performance evidenced by smaller DR and average DR values. However, the superiority in stabilizing traffic flow for the hybrid system is found deteriorating when the number of CAVs in a platoon system increases. This can be seen from the increased value of the DR in the hybrid system in Scenario 1. Under Scenario 2, all the DR and average DR become smaller for the three platoon systems. The results indicate that the stability performance of the platoon systems would be better in case of fewer oscillations resulted from the exogenous disturbance. Moreover, we observe that the hybrid system remains effective in guaranteeing stability through the whole system under large deceleration and acceleration conditions, even if it is slightly inferior to the CTG-based system.

Table 3. Dampening ratio of different platoon systems under Scenarios 1 and 2

| Scenario | System | $r$ | $n=4$ | $n=5$ | $n=6$ | $n=7$ | $n=8$ | $n=9$ | $n=10$ | Average |
|---|---|---|---|---|---|---|---|---|---|---|
| Scenario 1 | CS-based | 2 | 1.27 | 1.28 | 1.29 | 1.29 | 1.29 | 1.29 | 1.30 | 1.29 |
| | CTG-based | 2 | 0.96 | 0.96 | 0.95 | 0.94 | 0.94 | 0.93 | 0.93 | 0.95 |
| | | 3 | 0.96 | 0.96 | 0.95 | 0.95 | 0.94 | 0.94 | 0.93 | 0.95 |
| | Hybrid | 2 | 0.96 | 0.97 | 0.97 | 0.98 | 0.98 | 0.98 | 0.98 | 0.98 |
| Scenario 2 | CS-based | 2 | 1.16 | 1.17 | 1.17 | 1.17 | 1.17 | 1.17 | 1.17 | 1.17 |
| | CTG-based | 2 | 0.95 | 0.94 | 0.94 | 0.93 | 0.93 | 0.92 | 0.92 | 0.93 |
| | | 3 | 0.95 | 0.95 | 0.94 | 0.93 | 0.93 | 0.92 | 0.92 | 0.93 |
| | Hybrid | 2 | 0.95 | 0.95 | 0.95 | 0.95 | 0.95 | 0.95 | 0.95 | 0.95 |

*5.2.3. Energy and emissions*

In this subsection, we calculate the fuel consumption and emissions to evaluate the energy and emission performance of the three types of platoon systems based on the following VT-micro model (Ahn et al., 2002; Han et al., 2020):

$$In(MOE_i^k) = \sum_{c_1=0}^{3} \sum_{c_2=0}^{3} e_{c_1,c_2}^k v_i^{c_1} a_i^{c_2} \tag{34}$$

where $k$ denotes the category of energy and emission measurements, i.e., fuel consumption, HC, CO, $NO_x$, and $CO_2$, $MOE_i^k$ is the fuel consumption and emissions of vehicle $i$ at category $k$ (L/s or mg/s), $c_1$ and $c_2$ is the power of speed and acceleration ranging from 0 to 3, respectively, and $e_{c_1,c_2}^k$ is the regression coefficient, which is distinct for different categories under the speed power $c_1$ and acceleration power $c_2$. The values of $e_{c_1,c_2}^k$ used in the numerical experiments are adopted from Ahn et al. (2002). Table 4 summarizes the fuel consumptions and the emissions of the platoon systems under the aforementioned two scenarios. The results show that the hybrid platoon system has better performance in reducing fuel consumption and emissions than the CS-based and CTG-based platoon systems under Scenario 1. The improvements are attributed to the great synergy between the CTG and CS policies, which has effectively reduced the oscillatory magnitude of the acceleration and speed, thus leading to smaller fuel consumption and emissions. Of course, it is expected that the fuel consumptions and emissions are positively affected by the number of CAVs in the platoon system. Under Scenario 2, the hybrid system still performs consistently better in fuel consumption and emissions than the CS-based system. However, the results are mixed in the comparison with the CTG-based platoon system. The hybrid system consumes less fuel and emits less CO and $CO_2$, but unfavourably emits more HC and $NO_x$ than the CTG-based platoon system. This is probably due to the inferior asymptotic stability on acceleration of the hybrid system compared to the CTG-based system and the high sensitivities of the emissions of HC and $NO_x$ to disturbances under Scenario 2.

Table 4. Fuel and emissions of different platoon systems under Scenarios 1 and 2

| Scenario | Measurement | System | $r$ | $n=4$ | $n=5$ | $n=6$ | $n=7$ | $n=8$ | $n=9$ | $n=10$ | Average |
|---|---|---|---|---|---|---|---|---|---|---|---|
| Scenario 1 | Fuel | CS-based | 2 | 4.32 | 5.42 | 6.52 | 7.62 | 8.72 | 9.83 | 10.93 | 7.62 |
| | | CTG-based | 2 | 4.14 | 5.17 | 6.21 | 7.24 | 8.27 | 9.30 | 10.33 | 7.24 |
| | | | 3 | 4.14 | 5.18 | 6.21 | 7.24 | 8.27 | 9.31 | 10.34 | 7.24 |
| | | Hybrid | 2 | 4.10 | 5.13 | 6.17 | 7.21 | 8.25 | 9.28 | 10.32 | 7.21 |
| | HC | CS-based | 2 | 3.74 | 4.69 | 5.64 | 6.60 | 7.55 | 8.50 | 9.45 | 6.60 |
| | | CTG-based | 2 | 3.71 | 4.63 | 5.53 | 6.44 | 7.34 | 8.24 | 9.15 | 6.43 |
| | | | 3 | 3.71 | 4.63 | 5.54 | 6.44 | 7.34 | 8.24 | 9.14 | 6.43 |
| | | Hybrid | 2 | 3.66 | 4.58 | 5.50 | 6.42 | 7.35 | 8.27 | 9.20 | 6.43 |
| | CO | CS-based | 2 | 40.20 | 50.46 | 60.74 | 71.02 | 81.30 | 91.59 | 101.88 | 71.03 |
| | | CTG-based | 2 | 38.61 | 48.25 | 57.83 | 67.41 | 76.99 | 86.56 | 96.13 | 67.40 |
| | | | 3 | 38.62 | 48.26 | 57.84 | 67.41 | 76.98 | 86.54 | 96.10 | 67.39 |
| | | Hybrid | 2 | 38.19 | 47.83 | 57.48 | 67.13 | 76.78 | 86.43 | 96.08 | 67.13 |
| | $NO_x$ | CS-based | 2 | 4.26 | 5.43 | 6.62 | 7.81 | 9.00 | 10.19 | 11.38 | 7.81 |
| | | CTG-based | 2 | 3.92 | 4.88 | 5.66 | 6.42 | 7.19 | 7.95 | 8.70 | 6.39 |
| | | | 3 | 3.92 | 4.89 | 5.66 | 6.43 | 7.19 | 7.95 | 8.70 | 6.39 |
| | | Hybrid | 2 | 3.49 | 4.42 | 5.36 | 6.30 | 7.25 | 8.20 | 9.15 | 6.31 |
| | $CO_2$ | CS-based | 2 | 10,082 | 12,645 | 15,210 | 17,778 | 20,346 | 22,914 | 25,483 | 17,780 |
| | | CTG-based | 2 | 9,637 | 12,041 | 14,450 | 16,856 | 19,261 | 21,663 | 24,063 | 16,853 |
| | | | 3 | 9,640 | 12,046 | 14,454 | 16,860 | 19,264 | 21,666 | 24,066 | 16,857 |
| | | Hybrid | 2 | 9,539 | 11,951 | 14,364 | 16,778 | 19,193 | 21,608 | 24,023 | 16,779 |
| Scenario 2 | Fuel | CS-based | 2 | 4.05 | 5.07 | 6.10 | 7.12 | 8.14 | 9.16 | 10.19 | 7.12 |
| | | CTG-based | 2 | 4.00 | 5.00 | 6.00 | 7.00 | 7.99 | 8.99 | 9.99 | 6.99 |
| | | | 3 | 4.00 | 5.00 | 6.00 | 7.00 | 7.99 | 8.99 | 9.99 | 7.00 |
| | | Hybrid | 2 | 3.99 | 4.99 | 5.99 | 6.99 | 7.99 | 8.99 | 9.99 | 6.99 |
| | HC | CS-based | 2 | 3.95 | 4.94 | 5.93 | 6.93 | 7.92 | 8.91 | 9.90 | 6.93 |
| | | CTG-based | 2 | 3.79 | 4.73 | 5.70 | 6.66 | 7.64 | 8.60 | 9.57 | 6.67 |
| | | | 3 | 3.80 | 4.74 | 5.70 | 6.67 | 7.64 | 8.59 | 9.55 | 6.67 |
| | | Hybrid | 2 | 3.86 | 4.84 | 5.81 | 6.79 | 7.76 | 8.74 | 9.71 | 6.79 |
| | CO | CS-based | 2 | 44.02 | 55.12 | 66.22 | 77.34 | 88.45 | 99.57 | 110.68 | 77.34 |

| | System | $r$ | $n=4$ | $n=5$ | $n=6$ | $n=7$ | $n=8$ | $n=9$ | $n=10$ | Average |
|---|---|---|---|---|---|---|---|---|---|---|
| | CTG-based | 2 | 43.51 | 54.37 | 65.16 | 75.95 | 86.72 | 97.50 | 108.27 | 75.92 |
| | | 3 | 43.51 | 54.38 | 65.18 | 75.96 | 86.75 | 97.54 | 108.32 | 75.95 |
| | Hybrid | 2 | 43.21 | 54.05 | 64.89 | 75.73 | 86.57 | 97.42 | 108.26 | 75.73 |
| $NO_x$ | CS-based | 2 | 3.53 | 4.44 | 5.36 | 6.28 | 7.20 | 8.12 | 9.04 | 6.28 |
| | CTG-based | 2 | 1.79 | 2.15 | 2.95 | 3.56 | 4.35 | 4.97 | 5.75 | 3.64 |
| | | 3 | 1.88 | 2.22 | 2.89 | 3.65 | 4.31 | 4.88 | 5.46 | 3.61 |
| | Hybrid | 2 | 2.72 | 3.47 | 4.22 | 4.98 | 5.73 | 6.49 | 7.24 | 4.98 |
| $CO_2$ | CS-based | 2 | 9,360 | 11,716 | 14,073 | 16,431 | 18,790 | 21,149 | 23,509 | 16,432 |
| | CTG-based | 2 | 9,267 | 11,584 | 13,886 | 16,193 | 18,494 | 20,800 | 23,100 | 16,189 |
| | | 3 | 9,264 | 11,583 | 13,890 | 16,192 | 18,497 | 20,805 | 23,112 | 16,192 |
| | Hybrid | 2 | 9,221 | 11,529 | 13,837 | 16,146 | 18,456 | 20,765 | 23,075 | 16,147 |

Table 5. Surrogate safety measurements of different platoon systems under Scenarios 1 and 2

| Scenario | Measurement | System | $r$ | $n=4$ | $n=5$ | $n=6$ | $n=7$ | $n=8$ | $n=9$ | $n=10$ | Average |
|---|---|---|---|---|---|---|---|---|---|---|---|
| Scenario 1 | TET | CS-based | 2 | 0.00 | 0.00 | 0.00 | 0.00 | 0.00 | 0.00 | 0.00 | 0.00 |
| | | CTG-based | 2 | 0.00 | 0.00 | 0.00 | 0.00 | 0.00 | 0.00 | 0.00 | 0.00 |
| | | | 3 | 1.90 | 1.90 | 1.90 | 1.90 | 1.90 | 1.90 | 1.90 | 1.90 |
| | | Hybrid | 2 | 0.00 | 0.00 | 0.00 | 0.00 | 0.00 | 0.00 | 0.00 | 0.00 |
| | TIT | CS-based | 2 | 0.00 | 0.00 | 0.00 | 0.00 | 0.00 | 0.00 | 0.00 | 0.00 |
| | | CTG-based | 2 | 0.00 | 0.00 | 0.00 | 0.00 | 0.00 | 0.00 | 0.00 | 0.00 |
| | | | 3 | 0.08 | 0.08 | 0.08 | 0.08 | 0.08 | 0.08 | 0.08 | 0.08 |
| | | Hybrid | 2 | 0.00 | 0.00 | 0.00 | 0.00 | 0.00 | 0.00 | 0.00 | 0.00 |
| Scenario 2 | TET | CS-based | 2 | 0.00 | 0.00 | 0.00 | 0.00 | 0.00 | 0.00 | 0.00 | 0.00 |
| | | CTG-based | 2 | 0.00 | 0.00 | 0.00 | 0.00 | 0.00 | 0.00 | 0.00 | 0.00 |
| | | | 3 | 2.30 | 2.30 | 2.30 | 2.30 | 2.30 | 2.30 | 2.30 | 2.30 |
| | | Hybrid | 2 | 0.00 | 0.00 | 0.00 | 0.00 | 0.00 | 0.00 | 0.00 | 0.00 |
| | TIT | CS-based | 2 | 0.00 | 0.00 | 0.00 | 0.00 | 0.00 | 0.00 | 0.00 | 0.00 |
| | | CTG-based | 2 | 0.00 | 0.00 | 0.00 | 0.00 | 0.00 | 0.00 | 0.00 | 0.00 |
| | | | 3 | 0.49 | 0.49 | 0.49 | 0.49 | 0.49 | 0.49 | 0.49 | 0.49 |
| | | Hybrid | 2 | 0.00 | 0.00 | 0.00 | 0.00 | 0.00 | 0.00 | 0.00 | 0.00 |

*5.2.4. Safety*

This subsection aims to evaluate the safety performance of the three types of platoon systems in terms of two surrogate safety measurements named Time Exposed Time-to-collision (TET) and Time Integrated Time-to-collision (TIT). Both measurements are regarded as the modified surrogate safety measures of Time-to-collision (TTC) and have been widely used to evaluate safety performance (Cui et al., 2021; Zheng et al., 2020a). A larger TET or TIT value indicates a higher risk of a vehicle approaching oscillations. The detailed calculations are shown as follows:

$$TET = \sum_{t=1}^{T}\sum_{i=1}^{n} \gamma_t^i \Delta t, \text{ where } \gamma_t^i = \begin{cases} 1, \forall t \text{ s.t. } 0 < TTC_i(t) \le TTC^* \\ 0, \text{ otherwise} \end{cases} \tag{35}$$

$$TIT = \sum_{t=1}^{T} \sum_{i \in \{1,\dots,n | 0 < TTC_i(t) \le TTC^*\}} \left( \frac{1}{TTC_i(t)} - \frac{1}{TTC^*} \right) \Delta t \tag{36}$$

where the TTC of vehicle $i$ at time $t$, i.e. $TTC_i(t)$, is time that remains until a collision between the consecutive vehicles occurs if they maintain the current speeds when the subject vehicle $i$ move faster than the preceding vehicle ($i$-1), and it is set at infinity when the subject vehicle $i$ is slower than the preceding vehicle ($i$-1), and $TTC^*$ is the safety threshold, which is set at 2 s as suggested by Zheng et al. (2020a).

Table 5 presents all the (average) TET and TIT values of the platoon systems in the two scenarios. It can be seen that both the TET and TIT values of the hybrid platoon system, CS-based platoon system, and CTG-based platoon system ($r = 2$), are zero under Scenario 1, whereas the TET and TIT values of the CTG-based system ($r = 3$) are 1.90 and 0.08, respectively. The results suggest that the hybrid system and CS-based system have better performance in traffic safety as they have smaller TET or TIT values. This is because the hybrid (except for the leader) and CS-based systems aim to maintain the constant spacing, which is not affected by the driving speed. Therefore, the magnitude variation of the actual spacing is far smaller than the speed difference between the consecutive vehicles in the hybrid and CS-based systems. On the contrary, the CTG-based system exhibits a higher collision risk under Scenario 1, since the ratio of speed difference to the actual spacing can easily exceed the threshold value according to Eqs. (35) and (36) under the disturbances. Additionally, the CTG-based system ($r = 3$) with a smaller time gap has larger TET and TIT values than the CTG-based system ($r = 2$) using a larger time gap in Table 5. Therefore, it indicates that the CTG-based system ($r = 3$) has high collision risks under the disturbances, and the other platoon systems are not easily affected by the disturbances. Moreover, the safety performance appears insensitive to the number of CAVs in the platoon systems. Under Scenario 2, similar findings can be found that the safety performance of the CTG-based platoon system ($r = 3$) is still inferior to the other platoon systems. In conclusion, the findings demonstrate that both the hybrid and CS-based systems are superior to the CTG-based system in improving traffic safety.

### 5.2.5. Comfort

This subsection will present the comfort performance of the different platoon systems measured by the maximum jerk (Cui et al., 2021) calculated by

$$J_{max} = \max_{t \in [0,T]} |\dot{a}_i(t)| \tag{37}$$

The larger jerk the vehicle has, the less comfort it offers. The results are tabulated in Table 6. It shows that under Scenario 1, the CS-based platoon system is inferior to the other two types of platoon systems in terms of comfort. This is because the CS-based system fails to attenuate the acceleration, which results in the propagation of the exogenous disturbance and accordingly a larger jerk for the leader. On the contrary, the leading vehicle with a larger spacing in the hybrid and CTG-based systems can effectively mitigate the exogenous disturbance and thus produce a smaller jerk. Moreover, the mitigation ability of the hybrid system appears greater than the CTG-based system due to a larger time gap required for leader in the hybrid system. The comfort performance is insensitive to the number of CAVs in the platoon systems. Similar findings can also be observed under Scenario 2, in which the hybrid and CS-based systems have the best and worst performance, respectively. The difference between the hybrid system and the CTG-based system is eliminated. It indicates that the mitigation superiority of a larger time gap in the hybrid system is counteracted since stronger peak oscillations are produced in the platoon system under Scenario 2 (large deceleration and acceleration) than Scenario 1 (periodical fluctuation).

Table 6. Maximum jerk of different platoon systems under Scenarios 1 and 2

| Scenario | System | $r$ | $n = 4$ | $n = 5$ | $n = 6$ | $n = 7$ | $n = 8$ | $n = 9$ | $n = 10$ | Average |
|---|---|---|---|---|---|---|---|---|---|---|
| Scenario 1 | CS-based | 2 | 2.16 | 2.16 | 2.16 | 2.16 | 2.16 | 2.16 | 2.16 | 2.16 |
| | CTG-based | 2 | 1.57 | 1.57 | 1.57 | 1.57 | 1.57 | 1.57 | 1.57 | 1.57 |
| | | 3 | 1.57 | 1.57 | 1.57 | 1.57 | 1.57 | 1.57 | 1.57 | 1.57 |
| | Hybrid | 2 | 1.10 | 1.10 | 1.10 | 1.10 | 1.10 | 1.10 | 1.10 | 1.10 |
| Scenario 2 | CS-based | 2 | 5.00 | 5.00 | 5.00 | 5.00 | 5.00 | 5.00 | 5.00 | 5.00 |
| | CTG-based | 2 | 4.20 | 4.20 | 4.20 | 4.20 | 4.20 | 4.20 | 4.20 | 4.20 |

| | | 3 | 4.20 | 4.20 | 4.20 | 4.20 | 4.20 | 4.20 | 4.20 | 4.20 |
|---|---|---|------|------|------|------|------|------|------|------|
| Hybrid | | 2 | 4.20 | 4.20 | 4.20 | 4.20 | 4.20 | 4.20 | 4.20 | 4.20 |

### 5.3. Analysis of multi-platoon systems

To test the applicability of the hybrid platoon system on roads (i.e. CAV dedicated lanes), three types of multi-platoon systems composed of two single platoons of the same type, i.e., CS-based, CTG-based, and hybrid multi-platoon systems as shown in Figure 4 (a), (b), and (c), respectively, are considered for the analysis. The performance of multi-platoon systems in various aspects will be evaluated and compared. Unless stated otherwise, the parameter settings and scenarios in the numerical experiments are the same to the previous subsections.

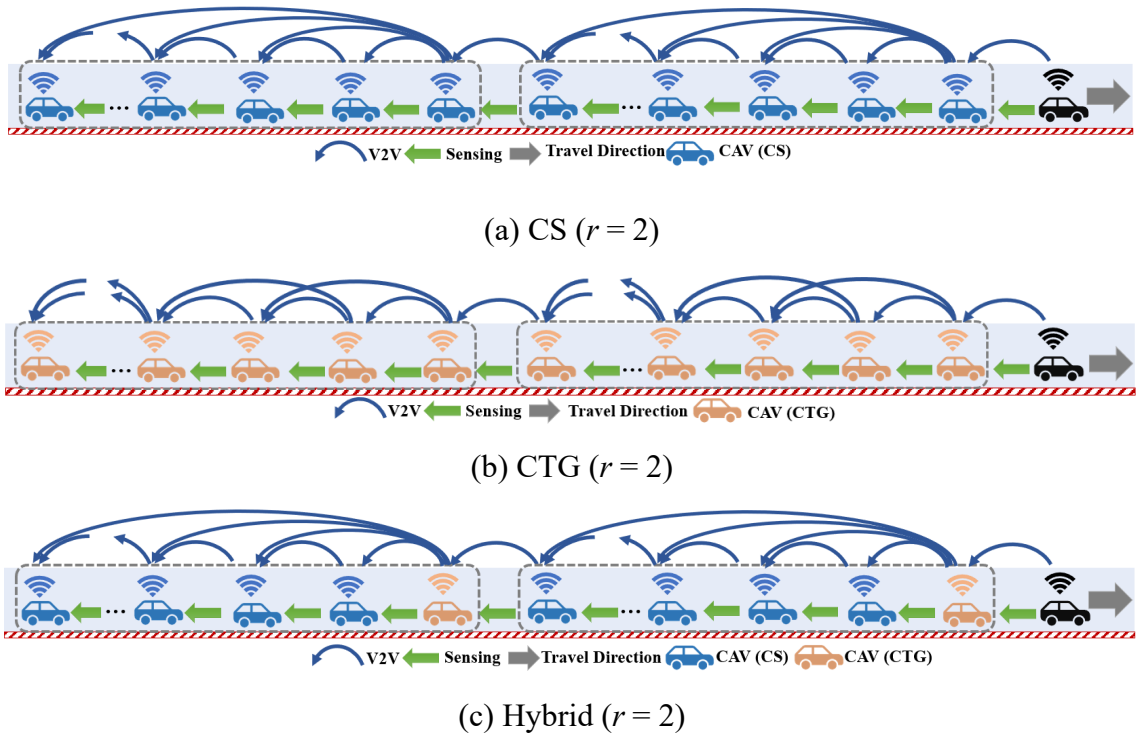

(a) CS ($r = 2$)

(b) CTG ($r = 2$)

(c) Hybrid ($r = 2$)

Figure 4. Illustration of multi-platoon systems with different spacing policies: (a) CS; (b) CTG; (c) Hybrid.

In this subsection, the performance of three types of multi-platoon systems with respect to efficiency, stability, fuel consumption, emissions, safety, and comfort will be compared. The results are summarized in Table 7. We can see that in terms of traffic efficiency, the hybrid multi-platoon systems are generally better than the CTG-based multi-platoon systems, even if the average traffic outflows of the CTG-based multi-platoon systems have slightly increased under Scenarios 1 and 2. The CS-based multi-platoon systems, without a doubt, have the best efficiency performance under any scenario. The findings further demonstrate that with more CAV members implementing the CS policy, the multi-platoon systems will reap more benefits in traffic efficiency.

As for the stability performance, the results confirm that the hybrid multi-platoon systems can effectively stabilize the traffic flows with the (average) dampening ratio value less than 1 under the two scenarios. It is further observed that the more platoon members the three types of multi-platoon systems have, the weaker stability performances against the exogenous disturbance they give under Scenario 2. Moreover, it is seen that the hybrid multi-platoon systems can still significantly reduce fuel consumption and emissions compared with the CS-based multi-platoon systems. However, similar to the comparative analysis of single platoon systems in Subsection 5.2.3, it has better performance in fuel consumption and the emissions of CO and $CO_2$ but inferior to the CTG-based multi-platoon systems for the emissions of HC and $NO_x$.

With respect to the safety performance, it shows that the TET or TIT and average TET or TIT values of CTG-based multi-platoon systems ($r = 3$) have increased upon the single CTG-based system ($r = 3$) (see Tables 5 and 7), and the TET or TIT values of the multi-platoon systems become smaller as the number of CAVs in the platoon increases under the two Scenarios. It suggests that the collision risks can be further increased in the CTG-based multi-platoon system when it incorporates more platoons but can be reduced if more CAVs are incorporated into the platoon. The safety performances of the hybrid and CS-based multi-platoon systems are not affected by the augmentation of platoons. They maintain better performance in traffic safety than the CTG-based multi-platoon systems with TET or TIT values being at zero thanks to the beneficial applications of CS policy. Furthermore, the hybrid multi-platoon systems show a visible advantage over the other multi-platoon systems in terms of comfort, especially in Scenario 1. It can further demonstrate that a larger time gap of the leader in each hybrid platoon system is helpful to dampen the maximum jerk of the CS-based systems in particular.

Table 7. MOEs of different multi-platoon systems under Scenarios 1 and 2

| MOE | Scenario | System | $r$ | $n=4$ | $n=5$ | $n=6$ | $n=7$ | $n=8$ | $n=9$ | $n=10$ | Average |
|---|---|---|---|---|---|---|---|---|---|---|---|
| $Q$ | Scenario 1 | CS-based | 2 | 2.28 | 2.32 | 2.31 | 2.29 | 2.32 | 2.31 | 2.32 | 2.30 |
| | | CTG-based | 2 | 1.29 | 1.31 | 1.33 | 1.34 | 1.35 | 1.36 | 1.38 | 1.34 |
| | | | 3 | 1.38 | 1.40 | 1.43 | 1.46 | 1.47 | 1.48 | 1.49 | 1.44 |
| | | Hybrid | 2 | 1.26 | 1.40 | 1.50 | 1.57 | 1.64 | 1.69 | 1.75 | 1.54 |
| | Scenario 2 | CS-based | 2 | 2.96 | 2.94 | 3.00 | 2.97 | 2.96 | 3.00 | 2.98 | 2.97 |
| | | CTG-based | 2 | 1.48 | 1.51 | 1.54 | 1.55 | 1.57 | 1.57 | 1.58 | 1.54 |
| | | | 3 | 1.60 | 1.64 | 1.69 | 1.71 | 1.74 | 1.75 | 1.75 | 1.70 |
| | | Hybrid | 2 | 1.48 | 1.64 | 1.76 | 1.89 | 1.97 | 2.04 | 2.13 | 1.84 |
| DR | Scenario 1 | CS-based | 2 | 1.27 | 1.29 | 1.29 | 1.30 | 1.30 | 1.30 | 1.30 | 1.30 |
| | | CTG-based | 2 | 0.97 | 0.96 | 0.95 | 0.95 | 0.94 | 0.94 | 0.93 | 0.95 |
| | | | 3 | 0.97 | 0.96 | 0.95 | 0.95 | 0.94 | 0.94 | 0.93 | 0.95 |
| | | Hybrid | 2 | 0.96 | 0.97 | 0.98 | 0.98 | 0.98 | 0.98 | 0.98 | 0.98 |
| | Scenario 2 | CS-based | 2 | 1.21 | 1.22 | 1.22 | 1.22 | 1.22 | 1.22 | 1.22 | 1.22 |
| | | CTG-based | 2 | 0.96 | 0.96 | 0.95 | 0.95 | 0.94 | 0.94 | 0.93 | 0.95 |
| | | | 3 | 0.96 | 0.96 | 0.95 | 0.95 | 0.94 | 0.94 | 0.93 | 0.95 |
| | | Hybrid | 2 | 0.96 | 0.97 | 0.97 | 0.97 | 0.97 | 0.97 | 0.97 | 0.97 |
| Fuel | Scenario 1 | CS-based | 2 | 8.96 | 11.26 | 13.57 | 15.88 | 18.18 | 20.48 | 22.79 | 15.87 |
| | | CTG-based | 2 | 8.27 | 10.33 | 12.39 | 14.44 | 16.49 | 18.54 | 20.59 | 14.44 |
| | | | 3 | 8.27 | 10.33 | 12.39 | 14.44 | 16.50 | 18.54 | 20.59 | 14.44 |
| | | Hybrid | 2 | 8.18 | 10.26 | 12.33 | 14.41 | 16.49 | 18.56 | 20.64 | 14.41 |
| | Scenario 2 | CS-based | 2 | 8.24 | 10.33 | 12.43 | 14.52 | 16.62 | 18.71 | 20.81 | 14.52 |
| | | CTG-based | 2 | 8.00 | 9.99 | 11.99 | 13.98 | 15.97 | 17.96 | 19.95 | 13.98 |
| | | | 3 | 8.00 | 10.00 | 11.99 | 13.98 | 15.98 | 17.97 | 19.96 | 13.98 |
| | | Hybrid | 2 | 7.98 | 9.98 | 11.98 | 13.98 | 15.99 | 17.99 | 19.99 | 13.99 |
| HC | Scenario 1 | CS-based | 2 | 7.75 | 9.74 | 11.74 | 13.74 | 15.74 | 17.74 | 19.73 | 13.74 |
| | | CTG-based | 2 | 7.40 | 9.25 | 11.01 | 12.79 | 14.57 | 16.37 | 18.17 | 12.79 |
| | | | 3 | 7.40 | 9.25 | 10.99 | 12.77 | 14.57 | 16.36 | 18.14 | 12.78 |
| | | Hybrid | 2 | 7.31 | 9.16 | 11.01 | 12.89 | 14.70 | 16.54 | 18.39 | 12.85 |

| | Scenario 2 | CS-based | 2 | 8.03 | 10.08 | 12.13 | 14.18 | 16.22 | 18.27 | 20.31 | 14.17 |
|---|---|---|---|---|---|---|---|---|---|---|---|
| | | CTG-based | 2 | 7.59 | 9.49 | 11.47 | 13.43 | 15.40 | 17.33 | 19.27 | 13.42 |
| | | | 3 | 7.57 | 9.47 | 11.44 | 13.37 | 15.29 | 17.23 | 19.14 | 13.36 |
| | | Hybrid | 2 | 7.74 | 9.68 | 11.63 | 13.58 | 15.53 | 17.47 | 19.42 | 13.58 |
| CO | Scenario 1 | CS-based | 2 | 84.89 | 107.12 | 129.36 | 151.57 | 173.75 | 195.92 | 218.08 | 151.53 |
| | | CTG-based | 2 | 77.15 | 96.35 | 115.40 | 134.45 | 153.52 | 172.61 | 191.71 | 134.46 |
| | | | 3 | 77.12 | 96.39 | 115.32 | 134.37 | 153.43 | 172.48 | 191.47 | 134.37 |
| | | Hybrid | 2 | 76.33 | 95.64 | 114.96 | 134.27 | 153.58 | 172.88 | 192.17 | 134.26 |
| | Scenario 2 | CS-based | 2 | 90.90 | 114.49 | 138.06 | 161.58 | 185.07 | 208.54 | 232.02 | 161.52 |
| | | CTG-based | 2 | 86.87 | 108.50 | 129.92 | 151.35 | 172.77 | 194.24 | 215.69 | 151.33 |
| | | | 3 | 86.89 | 108.57 | 129.94 | 151.41 | 172.92 | 194.38 | 215.82 | 151.42 |
| | | Hybrid | 2 | 86.32 | 107.99 | 129.67 | 151.35 | 173.03 | 194.70 | 216.38 | 151.35 |
| NO$_x$ | Scenario 1 | CS-based | 2 | 10.37 | 13.36 | 16.38 | 19.41 | 22.42 | 25.43 | 28.43 | 19.40 |
| | | CTG-based | 2 | 7.72 | 9.53 | 10.68 | 11.96 | 13.26 | 14.68 | 16.09 | 11.99 |
| | | | 3 | 7.69 | 9.57 | 10.52 | 11.87 | 13.25 | 14.65 | 15.89 | 11.92 |
| | | Hybrid | 2 | 6.86 | 8.74 | 10.60 | 12.54 | 14.42 | 16.30 | 18.15 | 12.52 |
| | Scenario 2 | CS-based | 2 | 7.76 | 9.89 | 12.03 | 14.17 | 16.30 | 18.43 | 20.55 | 14.16 |
| | | CTG-based | 2 | 3.71 | 4.76 | 6.63 | 8.41 | 10.25 | 11.60 | 13.13 | 8.36 |
| | | | 3 | 3.46 | 4.49 | 6.38 | 7.75 | 8.96 | 10.52 | 11.71 | 7.61 |
| | | Hybrid | 2 | 5.54 | 7.04 | 8.54 | 10.04 | 11.54 | 13.05 | 14.55 | 10.04 |
| CO$_2$ | Scenario 1 | CS-based | 2 | 20,882 | 26,254 | 31,625 | 36,994 | 42,359 | 47,723 | 53,085 | 36,989 |
| | | CTG-based | 2 | 19,239 | 24,022 | 28,832 | 33,621 | 38,407 | 43,178 | 47,943 | 33,583 |
| | | | 3 | 19,238 | 24,039 | 28,845 | 33,630 | 38,412 | 43,177 | 47,942 | 33,606 |
| | | Hybrid | 2 | 19,039 | 23,868 | 28,698 | 33,527 | 38,356 | 43,185 | 48,013 | 33,527 |
| | Scenario 2 | CS-based | 2 | 19,004 | 23,830 | 28,661 | 33,495 | 38,329 | 43,161 | 47,993 | 33,496 |
| | | CTG-based | 2 | 18,522 | 23,138 | 27,726 | 32,313 | 36,899 | 41,499 | 46,093 | 32,313 |
| | | | 3 | 18,530 | 23,153 | 27,737 | 32,339 | 36,946 | 41,539 | 46,143 | 32,341 |
| | | Hybrid | 2 | 18,432 | 23,051 | 27,672 | 32,293 | 36,915 | 41,537 | 46,159 | 32,294 |
| TET | Scenario 1 | CS-based | 2 | 0.00 | 0.00 | 0.00 | 0.00 | 0.00 | 0.00 | 0.00 | 0.00 |
| | | CTG-based | 2 | 0.00 | 0.00 | 0.00 | 0.00 | 0.00 | 0.00 | 0.00 | 0.00 |
| | | | 3 | 3.20 | 3.20 | 2.80 | 2.30 | 1.90 | 1.90 | 1.90 | 2.45 |

| | | | | | | | | | | |
|---|---|---|---|---|---|---|---|---|---|---|
| | | Hybrid | 2 | 0.00 | 0.00 | 0.00 | 0.00 | 0.00 | 0.00 | 0.00 | 0.00 |
| | Scenario 2 | CS-based | 2 | 0.00 | 0.00 | 0.00 | 0.00 | 0.00 | 0.00 | 0.00 | 0.00 |
| | | CTG-based | 2 | 0.00 | 0.00 | 0.00 | 0.00 | 0.00 | 0.00 | 0.00 | 0.00 |
| | | | 3 | 4.50 | 4.50 | 4.30 | 4.30 | 4.10 | 4.10 | 3.90 | 4.24 |
| | | Hybrid | 2 | 0.00 | 0.00 | 0.00 | 0.00 | 0.00 | 0.00 | 0.00 | 0.00 |
| TIT | Scenario 1 | CS-based | 2 | 0.00 | 0.00 | 0.00 | 0.00 | 0.00 | 0.00 | 0.00 | 0.00 |
| | | CTG-based | 2 | 0.00 | 0.00 | 0.00 | 0.00 | 0.00 | 0.00 | 0.00 | 0.00 |
| | | | 3 | 0.11 | 0.10 | 0.09 | 0.08 | 0.08 | 0.08 | 0.08 | 0.08 |
| | | Hybrid | 2 | 0.00 | 0.00 | 0.00 | 0.00 | 0.00 | 0.00 | 0.00 | 0.00 |
| | Scenario 2 | CS-based | 2 | 0.00 | 0.00 | 0.00 | 0.00 | 0.00 | 0.00 | 0.00 | 0.00 |
| | | CTG-based | 2 | 0.00 | 0.00 | 0.00 | 0.00 | 0.00 | 0.00 | 0.00 | 0.00 |
| | | | 3 | 0.75 | 0.75 | 0.70 | 0.66 | 0.64 | 0.61 | 0.59 | 0.67 |
| | | Hybrid | 2 | 0.00 | 0.00 | 0.00 | 0.00 | 0.00 | 0.00 | 0.00 | 0.00 |
| $J_{max}$ | Scenario 1 | CS-based | 2 | 2.89 | 2.85 | 2.89 | 2.89 | 2.89 | 2.89 | 2.89 | 2.89 |
| | | CTG-based | 2 | 1.56 | 1.56 | 1.56 | 1.56 | 1.56 | 1.56 | 1.56 | 1.56 |
| | | | 3 | 1.56 | 1.56 | 1.56 | 1.56 | 1.56 | 1.56 | 1.56 | 1.56 |
| | | Hybrid | 2 | 1.13 | 1.13 | 1.13 | 1.13 | 1.13 | 1.13 | 1.13 | 1.13 |
| | Scenario 2 | CS-based | 2 | 5.03 | 5.03 | 5.03 | 5.03 | 5.03 | 5.03 | 5.03 | 5.03 |
| | | CTG-based | 2 | 4.20 | 4.20 | 4.20 | 4.20 | 4.20 | 4.20 | 4.20 | 4.20 |
| | | | 3 | 4.20 | 4.20 | 4.20 | 4.20 | 4.20 | 4.20 | 4.20 | 4.20 |
| | | Hybrid | 2 | 4.20 | 4.20 | 4.20 | 4.20 | 4.20 | 4.20 | 4.20 | 4.20 |

## 6. Conclusions and Future Research

In this study, we propose a hybrid platoon system for connected and automated vehicles (CAVs) jointly using the constant time gap (CTG) and constant spacing (CS) policy based on the linear feedback and feedforward controllers. The hybrid spacing policy is firstly formulated, where the leader adopts the CTG policy and the followers use the CS policy in the platoon system. Based on the hybrid formulation, the $h_2$-norm string stability criteria related to spacing error and acceleration are used for the stability analysis. The notion of ex-head-to-tail string stability on acceleration for the platoon system is introduced and the sufficient stability conditions in the frequency domain are derived using the Laplace transform. The hybrid string stability of the hybrid platoon system combining the string stability conditions on spacing error and the ex-head-to-tail string stability conditions on acceleration are proposed. Moreover, numerical experiments are conducted to validate the mathematical proofs of stability analysis. Several measurements of effectiveness under two typical scenarios, i.e., periodical fluctuation and large deceleration and acceleration, are adopted to verify the effectiveness of the hybrid platoon system in the aspects of efficiency, stability, energy, emissions, safety, and comfort. The evaluation is further extended to multi-platoon systems composed of two single platoons as an example. The experimental results show that the hybrid (multi-) platoon systems perform better than the CS-based (multi-) platoon systems and CTG-based (multi-) platoon systems ($r$ = 2 and 3) in most aspects. Our study may serve as a useful guide for the hybrid platoon systems in practical application.

In a conclusion, some key results through the systematic analysis are summarized: (i) The hybrid platoon system can achieve hybrid string stability that guarantees the string stability on spacing error and ex-head-to-tail string stability on acceleration. The hybrid spacing policy can theoretically enhance the string stability of a platoon system compared to one using the CS policy only. It also provides a potential opportunity to implement the hybrid systems on the platoon level. (ii) The hybrid platoon system has much better performance than the CS-based platoon system in most aspects, including the stability, fuel consumption, emissions, safety, and comfort. (iii) The hybrid platoon system has better or competitive performance compared to CTG-based platoon system under the periodical fluctuation and large deceleration and acceleration scenarios. Essentially, it has neutralized the poor performance in terms of traffic efficiency and safety and the good performance regarding the stability of the CTG-based platoon system.

Further research work can be conducted in the following several aspects. First, other vehicle controllers with distinct formulations using these spacing policies can be further explored to make a comparison of the performance among the hybrid platoon system, CTG-based platoon system, and CS-based platoon system. Second, it would be interesting to design a variable time gap strategy between the CTG and CS policy for the hybrid system and further enhance the applicability of the system with guaranteed string stability. Third, the performance of the hybrid system is largely affected by the heterogeneity and uncertainty in vehicle dynamics and controller parameters, etc. It is thus necessary to propose more efficient and robust methods for future implementation of the hybrid system. In addition, the degradation and safety issues in case of communication failure would be also considered in future research.

## Appendix A. Notation

| | |
|---|---|
| $i$ | Vehicle longitudinal sequence |
| $n$ | Total number of the platoon system |
| $t, \tau$ | Time instant |
| $p$ | Front bumper position |
| $v$ | Speed |
| $a$ | Realized acceleration |
| $u$ | Desired acceleration |
| $r$ | Number of predecessors |
| $s$ | Actual spacing |
| $s^*$ | Target spacing |
| $\Delta s$ | Spacing error |
| $h$ | Constant time gap |
| $d$ | Inter-vehicle distance in standstill condition |
| $L$ | Vehicle length |
| $\delta$ | Sensor delay |
| $\theta$ | Communication delay |
| $\varphi$ | Actuation time lag |
| $g$ | Time delay |
| $\sigma$ | Accumulated time delay |
| $T$ | Simulation time |
| $\Delta t$ | Simulation time step |
| $k_s, k_v, k_a$ | Controller parameters of the leading vehicle |
| $q_1, q_3, q_4, \lambda$ | Controller parameters of the following vehicles |
| $Q$ | Traffic outflow |
| $DR$ | Dampening ratio |
| $TET$ | Time Exposed Time-to-collision |
| $TIT$ | Time Integrated Time-to-collision |
| $J_{max}$ | Maximum jerk |

## Appendix B. Proof of Proposition B.1

Proposition B.1: The hybrid platoon system is not head-to-tail string stable on acceleration.

**Proof.** By substituting the index ($i$-1) in Eq. (22), we have

$$
\begin{aligned}
A(z)p_{i-1}(z)e^{-g_iz} &= B(z)p_{i-2}(z)e^{-g_iz-g_{i-1}z} + C(z)p_1e^{-\sigma_{i-1}z-g_iz}\\
&= B(z)p_{i-2}(z)e^{-g_iz-g_{i-1}z} + C(z)p_1e^{-\sigma_iz}, \forall i \in \{3,...,n\}
\end{aligned}
\tag{B1}
$$

According to Eq. (B1), we substitute index ($i$-1) of the right-hand term in Eq. (23) as $i$ = 1 and obtain

$$
\begin{aligned}
p_i(z) &= \frac{B(z)}{A(z)}p_{i-1}(z)e^{-g_iz} + \frac{C(z)}{A(z)}p_1(z)e^{-\sigma_iz}\\
&= \frac{B(z)}{A(z)}\left[\frac{B(z)}{A(z)}p_{i-2}(z)e^{-g_iz-g_{i-1}z} + \frac{C(z)}{A(z)}p_1(z)e^{-\sigma_iz}\right] + \frac{C(z)}{A(z)}p_1(z)e^{-\sigma_iz}\\
&= (\frac{B(z)}{A(z)})^{i-1}p_1(z)e^{-g_iz...-g_2z} + (\frac{B(z)}{A(z)})^{i-2}\frac{C(z)}{A(z)}p_1(z)e^{-g_iz...-g_2z}... + \frac{B(z)}{A(z)}\frac{C(z)}{A(z)}p_1(z)e^{-\sigma_iz} + \frac{C(z)}{A(z)}p_1(z)e^{-\sigma_iz}\\
&= (\frac{B(z)}{A(z)})^{i-1}p_1(z)e^{-\sigma_iz} + \left[(\frac{B(z)}{A(z)})^{i-2}\frac{C(z)}{A(z)} + ... + \frac{B(z)}{A(z)}\frac{C(z)}{A(z)} + \frac{C(z)}{A(z)}\right]p_1(z)e^{-\sigma_iz}\\
&= (\frac{B(z)}{A(z)})^{i-1}p_1(z)e^{-\sigma_iz} + \frac{C(z)}{A(z)-B(z)}\left[1-(\frac{B(z)}{A(z)})^{i-1}\right]p_1(z)e^{-\sigma_iz}, \forall i \in \{2,...,n\}
\end{aligned}
\tag{B2}
$$

Then we can obtain the norm of the head-to-tail string stability on acceleration for the hybrid platoon system as follows:

$$
\begin{aligned}
\frac{\|a_n\|}{\|a_1\|} = \frac{\|p_n\|}{\|p_1\|} &= \left\|\frac{(\frac{B(z)}{A(z)})^{n-1}p_1(z)e^{-\sigma_nz} + \frac{C(z)}{A(z)-B(z)}\left[1-(\frac{B(z)}{A(z)})^{n-1}\right]p_1(z)e^{-\sigma_nz}}{p_1(z)}\right\|\\
&= \left\|(\frac{B(z)}{A(z)})^{n-1} + \frac{C(z)}{A(z)-B(z)}\left[1-(\frac{B(z)}{A(z)})^{n-1}\right]\right\|\|e^{-\sigma_nz}\|, \forall w > 0, z = jw
\end{aligned}
\tag{B3}
$$

The norm of the right-hand term on the Eq. (B3) is always larger than or equal to 1 for any controller parameters and index $n$ through numerical calculations. In other words, we cannot find the feasible controller parameters and index $n$ that the Eq. (17) in Definition 5 holds based on the local stability conditions (i.e. Eq. (11)). Therefore, we can conclude that the hybrid platoon system is not head-to-tail string stable on acceleration. This completes the Proposition B.1. □